\documentclass[12pt,thmsa]{article}

\usepackage{amsfonts}
\usepackage{graphicx}
\usepackage{epsfig}
\usepackage{graphics}

\makeatletter
\newcommand{\row}[1]%
{\mathord{\buildrel{\lower3pt%
\hbox{$\scriptscriptstyle\rightarrow$}}\over #1}}

\newcommand{\dyadic}[1]{\mathord{\dyadic@rrow{#1}}}
\newcommand{\dyadic@rrow}[1]{
\begin{picture}(12,12)(-1,0)
\put(-1,9){\makebox(0,0)[t]{$\scriptscriptstyle\downarrow$}}
\put(-1,9){\makebox(0,0)[l]{$\scriptscriptstyle\longrightarrow$}}
\put(5,0){\makebox(0,0)[b]{$#1$}}
\end{picture}
}

\newcommand{\bra}[1]{\bigl\langle #1 \bigr|}
\newcommand{\ket}[1]{\bigl| #1 \bigr\rangle}

\begin{document}

\begin{center}
{\Large Entanglement of simultaneous and non-simultaneous
Accelerated Qubit-Qutrit systems}\\
 {N. Metwally\\}
$^1$ Department of Mathematics, College of Science, Bahrain
University, Bahrain \\
$^2$Department of Mathematics, Faculty of Science,
Aswan University, Aswan, Egypt \\
email:nmetwally@gmail.com, nmetwally@uob.edu.bh \\
Tel:00973 33237474, ~ Fax: 00973 17449145
\end{center}
\date{\today }

\begin{abstract}
In this contribution, we investigate the entanglement behavior of
a composite system consists of two different dimensional
subsystems in non-inertial frames. In particular, we consider a
composite system of qubit(two-dimensional) subsystem, and qutrit,
(three-dimensional) subsystem. The degree of entanglement is
quantified for different cases, where it is assumed that the
two-subsystems are simultaneously or non- simultaneously
accelerated. The entanglement decays as the acceleration of any
subsystem increases.In general, the decay rate of entanglement
increases as  the dimensional of the accelerated subsystem
increases. These results could be  important in building an
accelerated quantum network consists of different dimensions
nodes.

\end{abstract}

\topmargin=0.001cm \textheight=24cm \textwidth=17cm
\section{Introduction}
It is well known that, quantum correlations are the most important
physical quantities that have many applications in the context of
quantum information, communication and computations
\cite{Nielsen,El1,El4}. The behavior of entanglement between
different systems   has investigated for small and high dimensions
systems \cite{E3}. However, the properties of  systems which are
consists of different dimensional  subsystems have investigated in
different ways. For example, the behavior of entanglement for a
two-parameter class of qubit-qutrit system passes through
dissipative channels   has discussed by Wei et.al, \cite{Wei}.
Karat and Gedik \cite{Karpat} have discussed the quantum and
classical correlation of qubit-qutrit systems in the presence of
classical dephasing environment. The  quantum correlation's
behavior  of a qubit-qutrit system under the effect of dephasing
and bit flip channels  has investigated  by Guo et. al.,
\cite{Long}. Moreover, X. Xiao \cite{Xiao} has investigated the
possibility of protecting the quantum correlation of qubit-qutrit
system by using weak measurement and reversal.

Recently, the behavior of entanglement  in non-inertial frames has
investigated for different  dimensional systems. For example,
Alsing et. al.,  \cite{Alsing1} have discussed  the decay of
entanglement between two modes of a free Dirac field in a
non-inertial frame. The  correlation between two modes of Dirac
fields in non-inertial frames have investigated by Wang et. al,
\cite{Wang}. The  dynamics of a general two qubit system in
non-inertial frame is investigated   by Metwally \cite{Metwally1}.
The possibility of using an accelerated class of  $X$-state to
perform quantum coding is discussed in \cite{Metwally2}. Moreover,
teleportation in non-inertial frame is discussed by several
authors see for example (\cite{Alsing, Alsing1,Metwally3}).

However, for composite system which consists of different
dimensions subsystems, there are some limited efforts have been
done to study the entangled  properties of these systems in
non-inertial frames. The effect of global, collective, local and
multilocal depolarizing  noise on qutrit-qutrit states is studied
by S. Khan and M. Khan \cite{khan}. The measurement induced
disturbance for qubit-qutrit system  in non-inertial frames under
the effect of different noise channel has  investigated by Hao and
Lian-Fu \cite{Hao}. The importance of this problem appears clearly
in the the context of generating accelerated network. To clarify
this task,  assume that a source  supplies randomly  some  users
who located in different nodes with different dimensional states
to generated entangled network \cite{metwally4}. So, may any two
nodes of the same or different dimensions systems are  connected.
Then it is important to quantify the degree of entanglement to
find the optimal connection. Therefor, we are motivated to
consider a general class of qubit-qutrit system in non-inertial
frame, where we assume that these subsystems simultaneous and
non-simultaneous accelerated.

This manuscript is organized as follows: In Sec. $2$, we define
the general form of the suggested system. The final state of the
system is obtained analytically when one or both subsystems are
accelerated  simultaneously or non- simultaneously. In Sec. $3$,
the survival of entanglement between the two subsystems is
quantified by using the negativity as  a measure  of entanglement.
Finally, the results are summarized in Sec. $4$.

\section{System and its evolution}
In this contribution, it is assumed that we have a composite
system, consists of qubit $(A)$  and qutrit $(B)$, is given in its
general form. The evolution the final state of the qubit-qutrit
($Q_{bt}$ ) system  is investigated for three different  cases:
(i) the subsystem $(A)$ (qubit) is accelerated while qutrit $(B)$
is in a rest,(ii) the subsystem $(B)$ (qutrit) is accelerated
while the subsystem $(A)$ is in a rest, (iii) Both subsystems  are
accelerated. The general  state of the qubit-qutrit  system is
given by,
\begin{equation}\label{Qbt}
\rho_{Q_{bt}}=\frac{1}{6}\Bigl\{I_{2}\otimes
I{_3}+\sum_{i=1}^{3}s_i\sigma_i\otimes I_3+
\sum_{i=1}^{8}t_j\tau_j\otimes
I_3+\sum_{i=1}^{3}\sum_{j=1}^{8}c_{ij}\sigma_i\otimes\tau_j\Bigr\},
\end{equation}
where $I_{2}$ and $I_{ 3}$ are  the unit matrix of size $2\times
2$ and $3\times 3$ respectively,
$s_i=tr\bigl\{\rho_{Q_{bt}}\sigma_i\otimes I_{3\time
3}\bigr\},\quad t_j=tr\bigl\{\rho_{Q_{bt}}I_{2\time2}\otimes
\tau_j\bigr\}$ are the Bloch vectors for the qubit and the qutrit
respectively, $i=1..3,\quad j=1..8$. The operators $\sigma_i$ and
$\tau_j$ are the traceless hermitian generators of $Su(2)$ and
$Su(3)$ respectively. The elements
$c_{ij}=tr\bigl\{\rho_{Q_{bt}}\sigma_i\otimes \tau_j\bigr\}$,
represent the correlation matrix  between the qubit $(A)$ and the
qutrit $(B)$. In the computational basis $0$, $1$ and $2$, the
operators $\sigma_i$ and $\tau_j$ can be written as,
\begin{eqnarray}
\sigma_1&=&\ket{0}\bra{1}+\ket{1}\bra{0}, \quad
\sigma_2=i\ket{0}\bra{1}-\ket{1}\bra{0}),\quad
\sigma_3=\ket{1}\bra{1}-\ket{0}\bra{0}, \nonumber\\
\tau_1&=&\ket{0}\bra{1}+\ket{1}\bra{0},\quad
\tau_2=i(\ket{1}\bra{0}-\ket{0}\bra{1}),\quad\tau_3=\ket{0}\bra{0}-\ket{1}\bra{1},
\nonumber\\
\tau_4&=&\ket{0}\bra{2}+\ket{2}\bra{0}, \quad
\tau_5=i(\ket{2}\bra{0}-\bra{0}\ket{2}),\quad
\tau_6=\ket{1}\bra{2}+\ket{2}\bra{1}, \nonumber\\
\tau_7&=&i(\ket{2}\bra{1}-\ket{1}\bra{2}),\quad
\tau_8=(\ket{0}\bra{0}+\ket{1}\bra{1}-2\ket{2}\bra{2})/\sqrt{3}.
\end{eqnarray}
In the computational  basis, $\ket{00},\ket{01}, \ket{02},
\ket{10}, \ket{11}$ and $\ket{12}$, the state (\ref{Qbt}) can be
written as,
\begin{eqnarray}
\varrho_{00,ij}&=&\mathcal{A}_{\ell_1},\quad  \ell_1=1...6, \quad
\quad  \varrho_{01,ij}=\mathcal{A}_{\ell_2}, \quad \ell_2=7...12,
\quad
 \nonumber\\
\varrho_{02,ij}&=&\mathcal{A}_{\ell_3} \quad \ell_3=13...18, \quad
\varrho_{10,ij}=\mathcal{A}_{\ell_4},\quad  \ell_4=19...24, \quad
\nonumber\\
 \varrho_{11,ij}&=&\mathcal{A}_{\ell_51},\quad
\ell_5=25...30, \quad \varrho_{12,ij}=\mathcal{A}_{\ell_6},\quad
\ell_6=31...36,\quad
\end{eqnarray}
where    $ij=00,01,02,10,11,12$ and $\mathcal{A}_{\ell},
\ell=1...36$  are given in the appendix (A). For example, the
element $\varrho_{00,00}=\mathcal{A}_1,\quad
\varrho_{00,01}=\mathcal{A}_2, \quad
\varrho_{02,02}=\mathcal{A}_{15}\quad
\varrho_{12,12}=\mathcal{A}_{36}$ and so on.

Now we investigate the evolution of the $Q_{bt}$ state (1) in
non-inertial frame when one or both of its subsystems are
accelerated as following:

\begin{enumerate}
\item{\it Only the qubit is accelertaed\\} In this context, it is
important to review the behavior of a particle in non-inertial
frames. It has been shown that, in the inertial frames, Minkowsik
coordinates $(t,z)$ are  used to describe Dirac field, while in
the uniformly accelerated case, Rindler coordinates $(\tau, x)$
are more adequate. The relations between the Minkowski and Rindler
coordinates  are given by\cite{Alsing,Edu},
\begin{equation}\label{trans}
\tau=r~tanh\left(\frac{t}{z}\right), \quad x=\sqrt{t^2-z^2},
\end{equation}
where  $-\infty<\tau<\infty$, $-\infty<x<\infty$  and $r$ is the
acceleration of the moving particle. The relations (\ref{trans})
describe  two regions in Rindler's spaces: the first region $I $
for $|t|<x$  and the second region $II$ for $x<-|t|$.

 A single mode $k$ of fermions and anti-fermions
in Minkowski space is described by the annihilation operators
$a_{kU}$  and $b_{-kU}$ respectively. In terms of Rindler's
operators ( $c^{(I)}_{kR}, d^{(II)}_{-kL}$), the Minkowski
operators can be written as \cite{Walls,un},
\begin{eqnarray}\label{op}
a_{kU}&=&\cos r c^{(I)}_{k,R}-\exp(-i\phi)\sin r d^{(II)}_{-k,L},
\nonumber\\
 b^\dagger_{-kU}&=&\exp(i\phi)\sin r
c^{(I)}_{k,R}+\cos r d^{(II)}_{k,L},
\end{eqnarray}
where, $tan~ r=e^{-\pi\omega \frac{c}{a}}$, $0\leq r\leq \pi/$4, $
a$ is the acceleration such that $0\leq a\leq\infty$, $\omega$ is
the frequency of the travelling qubits, $c$ is the speed of light,
and $\phi$ is an unimportant phase that can be absorbed into the
definition of the operators \cite{Jason2013}. The operators
(\ref{op}) mix a particle in region $I$ and an anti particle in
region $II$ as,
\begin{eqnarray}\label{Min}
\ket{0_k}&=&\cos r\ket{0_k}_I\ket{0_{-k}}_{II}+ \sin
r\ket{1_k}_I\ket{1_{-k}}_{II}, \nonumber\\
\ket{1_k}&=&a^\dagger_k\ket{0_k} =\ket{1_k}_I\ket{0_k}_{II}.
\end{eqnarray}
Now,  by using the initial state of the qubit-qutrit system
$\rho_{Q_{bt}}$  (\ref{Qbt})  and the operators (\ref{Min}), the
final state in the first region $I$ of the Rindler space can be
written in the basis $00,01, 02, 10, 11$ and $12$. The density
operator is represented by  a matrix of size $6\times 6$. Its
elements are given by,
\begin{eqnarray}
\rho_{00,00}^{(q)}&=&c^2_q\varrho_{00,00}, \quad
\rho_{00,01}^{(q)}=c^2_q\varrho_{00,01},\quad
\rho_{00,02}^{(q)}=c^2_q\varrho_{00,02},\quad
\rho_{00,10}^{(q)}=c_q\varrho_{00,10},
\nonumber\\
\quad \rho_{00,11}^{(q)}&=&c_q\varrho_{00,11},\quad
\rho_{00,12}^{(q)}=c_q\varrho_{00,12},\quad
\rho_{01,00}^{(q)}=c^2_q\varrho_{01,00}, \quad
\rho_{01,01}^{(q)}=c^2_q\varrho_{01,01},\quad
\nonumber\\
\rho_{01,02}^{(q)}&=&c^2_q\varrho_{01,02},\quad
\rho_{01,10}^{(q)}=c_q\varrho_{01,10},\quad
\rho_{01,11}^{(q)}=c_q\varrho_{01,11},\quad
\rho_{01,12}^{(q)}=c_q\varrho_{01,12},
\nonumber\\
\rho_{02,00}^{(q)}&=&c^2_q\varrho_{02,00}, \quad
\rho_{02,01}^{(q)}=c^2_q\varrho_{02,01},\quad
\rho_{02,02}^{(q)}=c^2_q\varrho_{02,02},\quad
\rho_{02,10}^{(q)}=c_q\varrho_{02,10},\quad
\nonumber\\
\rho_{02,11}^{(q)}&=&c_q\varrho_{02,11},\quad
\rho_{02,12}^{(q)}=c_q\varrho_{02,12},\quad
\rho_{10,00}^{(q)}=c_q\varrho_{10,00}, \quad
\rho_{10,01}^{(q)}=c_q\varrho_{10,01},\quad
\nonumber\\
\rho_{10,02}^{(q)}&=&c_q\varrho_{10,02},\quad
\rho_{10,10}^{(q)}=s^2_q\varrho_{00,00}+\varrho_{10,10},\quad
\rho_{10,11}^{(q)}=s^2_q\varrho_{00,01}+\varrho_{10,11},
\nonumber\\
\rho_{10,12}^{(q)}&=&s^2_q\varrho_{00,02}+\varrho_{10,12},\quad
\rho_{11,00}^{(q)}=c_q\varrho_{11,00}, \quad
\rho_{11,01}^{(q)}=c_q\varrho_{11,01},\quad
\rho_{11,02}^{(q)}=c_q\varrho_{11,02},\quad
\nonumber\\
\rho_{11,10}^{(q)}&=&s^2_q\varrho_{01,00}+\varrho_{11,10},\quad
\rho_{11,11}^{(q)}=s^2_q\varrho_{01,01}+\varrho_{11,11},\quad
\rho_{11,12}^{(q)}=s^2_q\varrho_{01,02}+\varrho_{11,12},
\nonumber\\
\rho_{12,00}^{(q)}&=&c_q\varrho_{12,00}, \quad
\rho_{12,01}^{(q)}=c_q\varrho_{12,01},\quad
\rho_{12,02}^{(q)}=c_q\varrho_{12,02},\quad
\rho_{12,10}^{(q)}=s^2_q\varrho_{02,00}+\varrho_{12,10},\quad
\nonumber\\
\rho_{12,11}^{(q)}&=&s^2_q\varrho_{02,01}+\varrho_{12,11},\quad
\rho_{12,12}^{(q)}=s^2_q\varrho_{02,02}+\varrho_{12,12},
\end{eqnarray}
where $c_q=\cos r_q$ and $s_q=sin r_q$.

\item{\it Only the qutrit is accelerated\\} In this case, it is
assumed  that the qubit is in an inertial frame, while the qutrit
undergoes a constant acceleration.  For the  two level  qubit
system (spin $1/2$) particles, the Pauli-principle allows only two
occupation numbers i.e.,"0" and "1" \cite{Alsing1}. However, the
possibility of allowing the occupation number $"2"$ is discussed
by Le$\grave{o}$n and Mart$\grave{i}$nez \cite{Juan}. Under the
single mode-approximation, Minlowaski vacuum state $\ket{0_M}$ in
the Rindler space reads as,
\begin{equation}\label{Va}
\ket{0_M}=cos^2r\ket{0}_{I}\ket{0}_{II}+e^{i\phi}\sin r\cos
r(\ket{\mathcal{U}}_{I}\ket{\mathcal{D}}_{II}+\ket{\mathcal{D}}_{I}\ket{\mathcal{U}}_{II})
+e^{2i\phi}sin^2r\ket{\mathcal{D}}_{I}\ket{\mathcal{P}}_{II},
\end{equation}
where $\ket{\mathcal{U}}, \ket{\mathcal{D}}$ and
$\ket{\mathcal{P}}$ are the spin up, spin down and pair state
respectively. The spin up and down  in Rindler space are given by,
\begin{eqnarray}\label{qt}
\ket{\mathcal{U}_M}&=&cos
r\ket{\mathcal{U}}_I\ket{0}_{II}+e^{i\phi}\sin
r\ket{\mathcal{P}}_I\ket{\mathcal{U}}_{II},
\nonumber\\
\ket{\mathcal{D}_M}&=&cos
r\ket{\mathcal{D}}_I\ket{0}_{II}-e^{i\phi}\sin
r\ket{\mathcal{P}}_I\ket{\mathcal{D}}_{II}.
\end{eqnarray}
By using the initial state (\ref{Qbt}) and the transformation
(\ref{Va},\ref{qt}) one can obtain the final  state of the
accelerated system. After tracing out the mode in the second
region $II$, we obtain the final accelerated state in the first
region  $I$ of the Rindler space. In the  basis,
$\ket{\ell,0},\ket{\ell,\mathcal{D}},\ket{\ell,\mathcal{U}},
\ket{\ell,\mathcal{P}}~( \ell=0,1)$, the density operator is
defined by a matrix of size $8\times 8$, where its elements are
given by,
\begin{eqnarray}
 \rho^{(t)}_{00,00}&=&c_t^4\varrho_{00,00},
 \quad \rho{(t)}_{00,0\mathcal{D}}=c_t^3\varrho_{00,01}
 ,\quad \rho{(t)}_{00,0\mathcal{U}}=c_t^2\varrho_{00,02} , \quad
 \rho{(t)}_{00,0\mathcal{P}}=0,\quad
\nonumber\\
\rho{(t)}_{00,10}&=&c_t^4\varrho_{00,10},\quad
\rho^{(t)}_{00,1\mathcal{D}}=c_t^3\varrho_{00,11},\quad
\rho^{(t)}_{00,1\mathcal{U}}=c_t^2\varrho_{00,12},
 \quad \rho_{00,1\mathcal{P}}=0,\quad
\rho^{(t)}_{0\mathcal{D},00}=c_t^4\varrho_{01,00},\quad
 \nonumber\\
\rho^{(t)}_{0\mathcal{D},0\mathcal{D}}&=&c_t^2(s^2_t\varrho_{00,00}+\varrho_{01,01}),\quad
 \rho^{(t)}_{0\mathcal{D},0\mathcal{U}}=c_t^2\varrho_{01,02},
  \quad \rho_{0\mathcal{D},0\mathcal{P}}=c_ts^2_t\varrho_{00,02},
  \nonumber\\
\rho^{(t)}_{0\mathcal{D},\mathcal{D}0}&=&c_t^4\varrho_{01,10},\quad
\rho^{(t)}_{0\mathcal{D},\mathcal{D}\mathcal{D}}=c^2_t(s^2_t\varrho_{00,10}+\varrho_{01,11}),\quad
\rho^{(t)}_{0\mathcal{D},1\mathcal{U}}=c_t^2\varrho_{01,12},\quad
\rho^{(t)}_{0\mathcal{D},1\mathcal{P}}=c_ts^2_t\varrho_{00,12},
 \nonumber\\
\rho^{(t)}_{0\mathcal{U},00}&=&c_t^3\varrho_{02,00},\quad
\rho^{(t)}_{02\mathcal{U},0\mathcal{D}}=c_t^3\varrho_{02,01},\quad
\rho^{(t)}_{0\mathcal{U},0\mathcal{U}}=c^2_t(s^2_t\varrho_{00,01}+\varrho_{02,02}),\quad
\rho^{(t)}_{0\mathcal{U},0\mathcal{P}}=-c_ts^2_t\varrho_{02,10},\quad
\nonumber\\
\rho^{(t)}_{0\mathcal{U},10}&=&c_t^3\varrho_{02,10},\quad
\rho^{(t)}_{0\mathcal{U},1\mathcal{D}}=c_t^2\varrho_{02,11},\quad
\rho^{(t)}_{0\mathcal{U},1\mathcal{U}}=c^2_t(s^2_t\varrho_{02,10}+\varrho_{02,12}),\quad
\rho^{(t)}_{0\mathcal{U},1\mathcal{P}}=-c_ts^2_t\varrho_{00,11},\quad
 \nonumber\\
\rho^{(t)}_{0\mathcal{P},00}&=&0,\quad
\rho^{(t)}_{0\mathcal{P},0\mathcal{D}}=c_ts_t^2\varrho_{02,00},\quad
\rho^{(t)}_{0\mathcal{P},0\mathcal{U}}=-c_ts_t^2\varrho_{01,00},\quad
\rho^{(t)}_{0\mathcal{P},0\mathcal{P}}=s^2_t(s^2_t\varrho_{00,00}+\varrho_{01,02}),\quad
\nonumber\\
\rho^{(t)}_{0\mathcal{P},10}&=&0,\quad
 \rho^{(t)}_{0\mathcal{P},1\mathcal{D}}=c_ts_t^2\varrho_{02,10},\quad
\rho^{(t)}_{0\mathcal{P},1\mathcal{U}}=-c_ts_t^2\varrho_{01,10},\quad
\rho^{(t)}_{0\mathcal{P},1\mathcal{P}}=s_t^4\varrho_{00,10},
\nonumber\\
\rho^{(t)}_{10,00}&=&c^4_t\varrho_{10,00},\quad
\rho^{(t)}_{10,0\mathcal{D}}=c^3_t\varrho_{10,01},\quad
\rho^{(t)}_{10,0\mathcal{U}}=c^2_t\varrho_{10,02},\quad
\rho^{(t)}_{10,0p}=0
\nonumber\\
\rho^{(t)}_{10,10}&=&c^4_t\varrho_{10,10},\quad
\rho^{(t)}_{10,1\mathcal{D}}=c^3_t\varrho_{10,11},\quad
\rho^{(t)}_{10,12\mathcal{U}}=c^2_t\varrho_{10,12}, \quad
\rho^{(t)}_{10,1\mathcal{P}}=0,
 \nonumber\\
\rho^{(t)}_{1\mathcal{D},00}&=&c^3_t\varrho_{11,00},\quad
\rho^{(t)}_{1\mathcal{D},0\mathcal{D}}=c^2_t\varrho_{11,01},\quad
\rho^{(t)}_{1\mathcal{D},0\mathcal{U}}=c^2_t\varrho_{11,02}, \quad
\rho^{(t)}_{1\mathcal{D},0\mathcal{P}}=c_ts^2_t\varrho_{10,02},
\nonumber\\
\rho^{(t)}_{1\mathcal{D},10}&=&c^3_t\varrho_{11,10},\quad
\rho^{(t)}_{1\mathcal{D},1\mathcal{D}}=c^2_ts^2_t\varrho_{11,11},\quad
\rho^{(t)}_{1\mathcal{D},1\mathcal{U}}=c^2_t\varrho_{11,12}, \quad
 \rho^{(t)}_{1\mathcal{D},1\mathcal{P}}=c^2_ts^2_t\varrho_{10,12},
 \nonumber\\
\rho^{(t)}_{1\mathcal{U},00}&=&c^2_t\varrho_{12,00},\quad
\rho^{(t)}_{1\mathcal{U},0\mathcal{D}}=c^2_t\varrho_{12,01},\quad
\rho^{(t)}_{1\mathcal{U},0\mathcal{U}}=c^2_t\varrho_{12,02}, \quad
\rho^{(t)}_{1\mathcal{U},0\mathcal{P}}=-c_ts^2_t\varrho_{10,01},
\nonumber\\
\rho^{(t)}_{1\mathcal{U},10}&=&c^2_t\varrho_{12,10},\quad
\rho^{(t)}_{1\mathcal{U},1\mathcal{D}}=c^2_t\varrho_{12,11},\quad
\rho^{(t)}_{1\mathcal{U},1\mathcal{U}}=c^2_ts^2_t\varrho_{10,10},
\quad
 \rho^{(t)}_{1\mathcal{U},1\mathcal{P}}=-c^2_ts_t\varrho_{10,112},
 \nonumber\\
\rho^{(t)}_{1\mathcal{P},00}&=&0,\quad
\rho^{(t)}_{1\mathcal{P},0\mathcal{D}}=c^2_ts^2_t\varrho_{12,00},\quad
\rho^{(t)}_{1\mathcal{P},0\mathcal{U}}=-c_ts^2_t\varrho_{11,00},
\quad
\rho^{(t)}_{1\mathcal{P},0\mathcal{P}}=s^2_t(\varrho_{11,01}+\varrho_{12,02}),
\nonumber\\
\rho^{(t)}_{1\mathcal{P},10}&=&0,
 \rho^{(t)}_{1\mathcal{P},1\mathcal{D}}=c^2_t\varrho_{12,11},\quad
\rho^{(t)}_{1\mathcal{P},1\mathcal{U}}=-c_ts^2_t\varrho_{11,10},
\quad
 \rho^{(t)}_{1\mathcal{P},1\mathcal{P}}=s^2_t(\varrho_{11,12}+\varrho_{12,12}),
\end{eqnarray}
where $c_t=\cos r_t$ and $s_t=sin~ r_t$.

 \item{\it Both subsystems are accelerated:\\}
 In this case, we assume that both subsystems are accelerated. The  final state of the system in
the first region $I$ of the Rindler space is defined by a matrix
of size $8\times 8$ elements. The elements of this matrix are
given by,
\begin{eqnarray}
\rho^{(qt)}_{00,jk}&=&\mathcal{B}^{(qt)}_{\ell_1},\quad
\ell_1=1..3,\quad jk=00,0\mathcal{D},0\mathcal{U},\quad
\rho^{(qt)}_{00,jk}=\mathcal{B}_{\ell_2},\quad
\ell_2=4,5,\quad jk=10,1\mathcal{D}, \nonumber\\
 \rho^{(qt)}_{00,12}&=&\varrho_{00,1\mathcal{P}}=0, \quad
 \rho_{01,jk}^{(qt)}=\mathcal{B}_{\ell_3}, \quad
 \ell_3=\mathcal{B}_6...\mathcal{B}_{13}, \quad
 jk=00,0\mathcal{D},0\mathcal{U},0\mathcal{P},10,1\mathcal{D},1\mathcal{U},1\mathcal{P},\quad
 \nonumber\\
 \rho^{(qt)}_{02,jk}&=&\mathcal{B}_{\ell_4}, \quad
 \ell_4=\mathcal{B}_{14}...\mathcal{B}_{21},\quad
 jk=00,0\mathcal{D},0\mathcal{U},0\mathcal{P},10,1\mathcal{D},1\mathcal{U},1\mathcal{P},\quad
\rho^{(qt)}_{0\mathcal{P},00}= \rho_{0\mathcal{P},10}=0,\quad
\nonumber\\
\varrho^{(qt)}_{0\mathcal{P},jk}&=&\mathcal{B}_{\ell_5}, \quad
 \ell_5=\mathcal{B}_{22}...\mathcal{B}_{27},\quad
jk=0\mathcal{D},0\mathcal{U},0\mathcal{P},1\mathcal{D},1\mathcal{U},1\mathcal{P}
 \quad
\rho^{(qt)}_{10,0\mathcal{P}}= \rho^{(qt)}_{10,1\mathcal{P}}=0,
\nonumber\\
\rho^{(qt)}_{10,jk}&=&\mathcal{B}_{\ell_6}, \quad
 \ell_6=\mathcal{B}_{28}...\mathcal{B}_{33},\quad
 jk=00,0\mathcal{D},0\mathcal{U},10,1\mathcal{D},1\mathcal{U},\quad
 \rho_{11,00}=0,\quad
 \nonumber\\
\rho^{(qt)}_{1\mathcal{D},jk}&=&\mathcal{B}_{\ell_7}, \quad
 \ell_7=\mathcal{B}_{34}...\mathcal{B}_{40},\quad
 jk=0\mathcal{D}, 0\mathcal{U},0\mathcal{P},10,1\mathcal{D},1\mathcal{U},1\mathcal{P},
 \nonumber\\
\rho^{(qt)}_{1\mathcal{U},jk}&=&\mathcal{B}_{\ell_8}, \quad
 \ell_8=\mathcal{B}_{41}...\mathcal{B}_{48},\quad
 jk=00,0\mathcal{D},0\mathcal{U},10,1\mathcal{D},1\mathcal{U},1\mathcal{P},
 \nonumber\\
 \rho^{(qt)}_{1\mathcal{P},00}&=&
\rho^{(qt)}_{1\mathcal{P},0\mathcal{U}}=\rho^{(qt)}_{1\mathcal{P},10}=0,\quad
\rho^{(qt)}_{1\mathcal{P},0\mathcal{D}}=\mathcal{B}_{48},\quad
\rho^{(qt)}_{1\mathcal{P},0\mathcal{P}}=\mathcal{B}_{49},\quad
\nonumber\\
\rho^{(qt)}_{1\mathcal{P},jk}&=&\mathcal{B}_{\ell_9}, \quad
\ell_9=50...52, \quad jk=1\mathcal{D},1\mathcal{U},1\mathcal{P},
\end{eqnarray}
where $\mathcal{B}_i$ are given in the appendix (B). For example
$\rho^{(qt)}_{00,00}=\mathcal{B}^{(qt)}_1$,\quad
$\rho^{(qt)}_{10,0\mathcal{D}}=\mathcal{B}^{(qt)}_{29}$,
$\rho^{(qt)}_{1\mathcal{P},1\mathcal{P}}=\mathcal{B}^{(qt)}_{52}$
and so on.

\end{enumerate}

\section{Dynamics of Entanglement}
To quantify the degree of entanglement $\mathcal{E}$ which is
contained in the accelerated  system, we use the negativity as a
measure. For any composite system $\rho_{ab}$ consists of two
subsystems with different dimensions. In our case, the system $a$
represents the qubit (2-dimensions) and $b$ refers to the qutrit
(3-dimensions). The negativity  for this system is defined as,
\begin{equation}
{\Large\mathcal{E}}=max\left(0,\sum_i{\lambda_i}\right),
\end{equation}
where $\lambda_i,i=1..6$ are the eigenvalues of the partial
transpose of $\rho^{T_a}_{ab}$ \cite{karpat}.

\subsection{Example one}

To investigate the effect of the accelerated  particles on the
degree of entanglement, we consider  a system of qubit-qutrit is
defined by,
\begin{eqnarray}
\row{s}&=&(0,0,s_3), \quad \row{t}=(0,0,t_3,0,0,0,0,0),
\nonumber\\
&&\mbox{and the non zero elements of the correlation matrix  is, }
%between the two-subsystems are given by}
\nonumber\\
 c_{11}&=&1,\quad c_{22}=-1, \quad c_{33}=1.
\end{eqnarray}
By using (7), we obtain the final state of the qubit-qutrit system
in the first region, $I$ of the Rindler space  when  only the
qubit is accelerated. This state is described by $6\times 6$
elements in the computational basis $00,01,02,10,11$ and $12$. The
non- zero elements of this accelerated state in region $I$ are
given by,
\begin{eqnarray}
\rho^{(q)}_{00,00}&=&c_q^2\varrho_{00,00},\quad
\rho^{(q)}_{00,11}=c_q\varrho_{00,11}, \quad
\rho^{(q)}_{01,01}=c_q^2\varrho_{01,01}, \quad
\rho^{(q)}_{01,10}=c_q\varrho_{01,10}
\nonumber\\
\rho^{(q)}_{02,02}&=&c_q^2\varrho_{02,02},\quad
\rho^{(q)}_{10,01}=c\varrho_{00,01}, \quad
\rho^{(q)}_{10,10}=\varrho_{10,10}+s_q^2\varrho_{00,00},
 \nonumber\\
\rho^{(q)}_{11,00}&=&c_q\varrho_{11,00},\quad
\rho^{(q)}_{11,11}=\varrho_{11,11}+s_q^2\varrho_{01,01}, \quad
\rho^{(q)}_{12,12}=\varrho_{12,12}+s_q^2\varrho_{01,02}.
\end{eqnarray}
Similarly,  If  only the qutrit is accelerated, then the non zero
elements of the  final state  of the accelerated  system in the
first region of Rindler space  can be written in  the basis
$\ket{\ell,0}, \ket{\ell,\mathcal{D}}, \ket{\ell,\mathcal{U}},
\ket{\ell,\mathcal{P}}$, ( $\ell=0,1)$  as,
\begin{eqnarray}
\rho^{(t)}_{00,00}&=&c_t^4\varrho_{00,00},\quad
\rho^{(t)}_{00,1\mathcal{D}}=c_t^3\varrho_{00,11}, \quad
\rho^{(t)}_{0\mathcal{D},0\mathcal{D}}=c_t^2\varrho_{01,01}+c^2_ts^2_t,
\quad \rho^{(t)}_{0\mathcal{D},10}=c_t^3\varrho_{01,10},
 \nonumber\\
\rho^{(t)}_{0\mathcal{P},0\mathcal{P}}&=&s_t^4\varrho_{00,00}+s^2_t(\varrho_{01,01}+\varrho_{02,02}),
\quad
\rho^{(t)}_{0\mathcal{P}p,1\mathcal{U}}=-c_ts^2_t\varrho_{01,10},\quad\rho^{(t)}_{02,1p}=-c^2_ts^2_t\varrho_{00,11},
\nonumber\\
\rho^{(t)}_{0\mathcal{U},0\mathcal{U}}&=&c_t^2s^2_t\varrho_{00,00}+c^2_t\varrho_{02,02},\quad
\rho^{(t)}_{10,0\mathcal{D}}=c^3_t\varrho_{10,01}, \quad
\rho^{(t)}_{10,10}=c^4_t\varrho_{10,10},
 \nonumber\\
\rho^{(t)}_{1\mathcal{D},00}&=&c_t^2\varrho_{11,00},\quad
\rho^{(t)}_{1\mathcal{D},1\mathcal{D}}=c^2_t\varrho_{11,11}+c_t^2s\varrho_{10,10},
\quad
\rho^{(t)}_{1\mathcal{U},1\mathcal{U}}=c^2_t\varrho_{12,12}+c^2_ts_t^2\varrho_{10,10},
\nonumber\\
\rho^{(t)}_{1\mathcal{U},0\mathcal{P}}&=&-c_ts^2_t\varrho_{10,01},\quad
\rho^{(t)}_{1\mathcal{P},0\mathcal{U}}=-c_ts^2_t\varrho_{11,00},
\quad\rho^{(t)}_{1\mathcal{P},1\mathcal{P}}=s_t^4\varrho_{10,10}+s_t^2(\varrho_{11,11}+\varrho_{12,12}).
\nonumber\\
\end{eqnarray}
\begin{figure}
  \begin{center}
     \includegraphics[width=19pc,height=15pc]{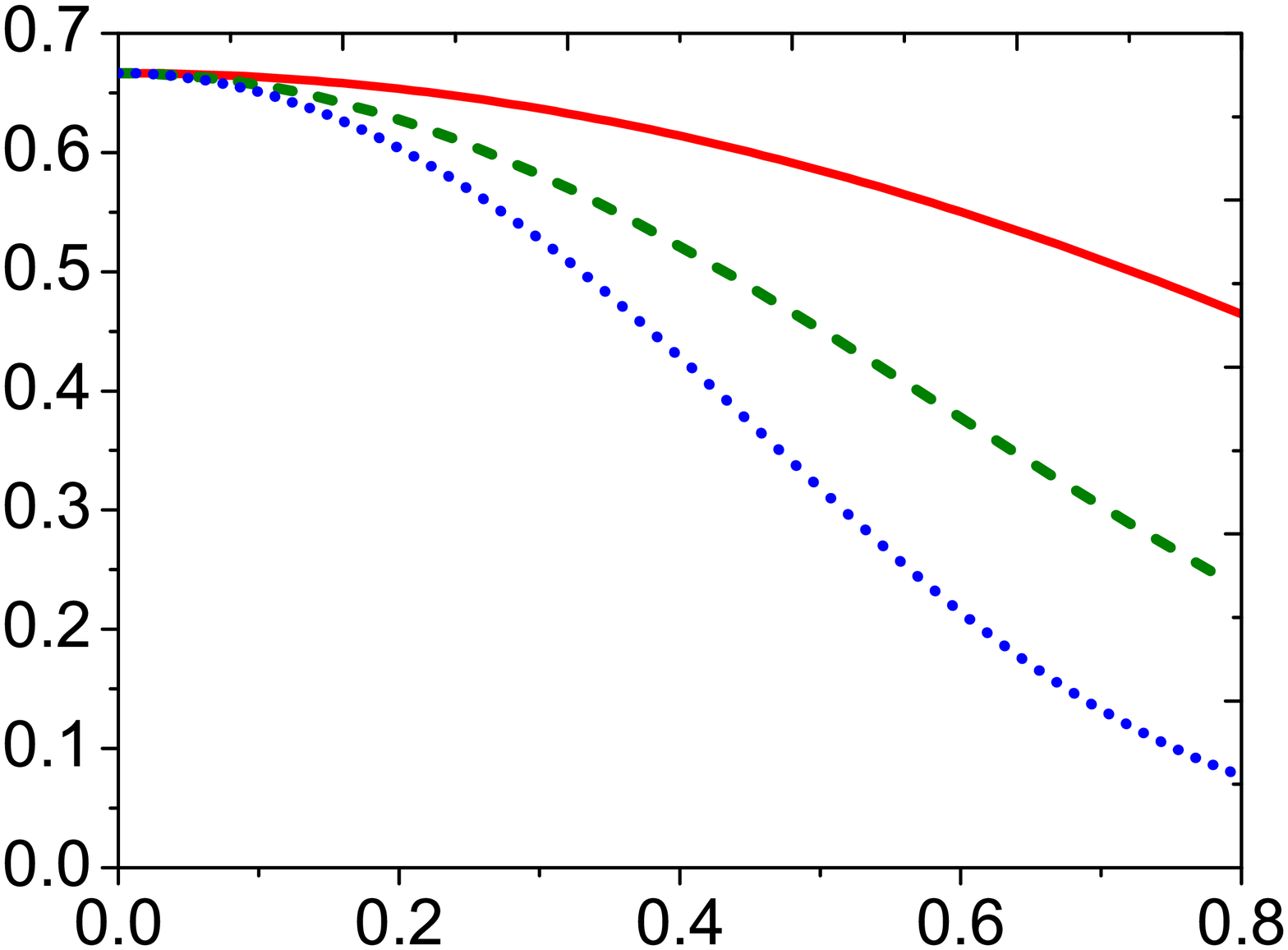}
 \includegraphics[width=19pc,height=15pc]{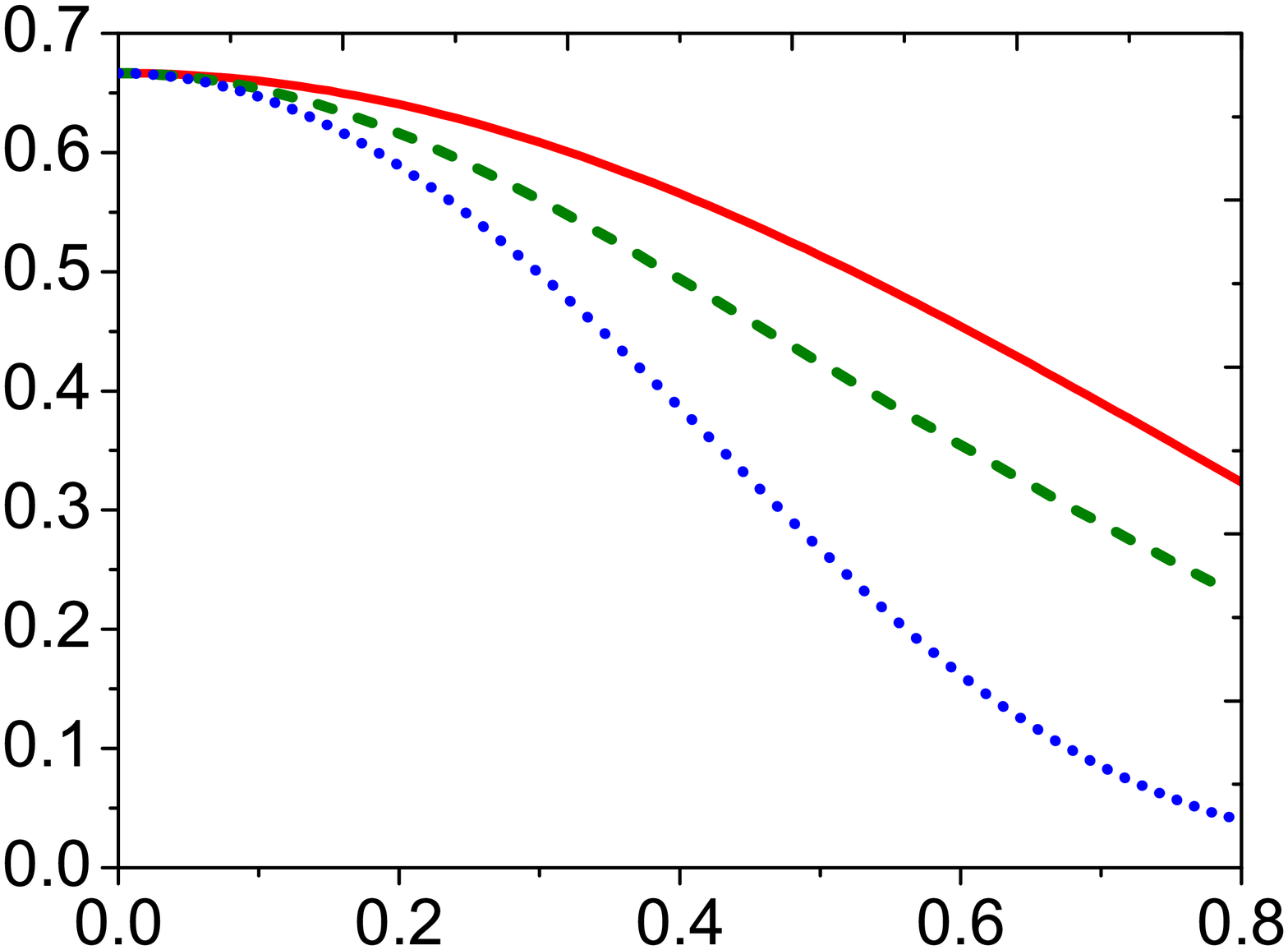}
  \put(-100,1){$r$}
  \put(-450,90){\Large$\mathcal{E}$}
   \put(-350,1){$r$}
     \put(-220,90){\Large$\mathcal{E}$}
      \caption{The degree of entanglement  $\mathcal{E}$, where we assume that $r_q=r_t=r$ and (a)
     $ s_3=1$, $t_3=1$, and  (b)$s_3=t_3=0$. The solid  and the dot lines shows the behavior of entanglement
       when only the qubit and only the qutrit are accelerated respectively.   }
  \end{center}
\end{figure}
Finally, let us assume that both particles are accelerated. In
this case, the  elements of the final state in the first region
$I_1$ are given by,
\begin{eqnarray}
\rho^{(qt)}_{00,00}&=&c_t^4c^2_q\varrho_{00,00},\quad
\rho^{(qt)}_{0\mathcal{D},0\mathcal{D}}=c^2_qc^2_t(\varrho_{01,01}+s_t^2),
\quad\rho^{(qt)}_{10,01}=c_qc_t^3(\varrho_{10,01}+\varrho_{02,01}),
\nonumber\\
\rho^{(qt)}_{0\mathcal{U},0\mathcal{U}}&=&c^2_qc^2_t(\varrho_{02,02}+\varrho_{00,00}s^2_t),
\quad\rho^{(qt)}_{1\mathcal{P},0\mathcal{U}}=-c_qc_ts^2_t\varrho_{11,00},\quad
\nonumber\\
\rho^{(qt)}_{0\mathcal{P},0\mathcal{P}}&=&c^2_qs^2_t(s^2_q\varrho_{00,00}+\varrho_{01,01}+\varrho_{02,02}),\quad
\rho^{(qt)}_{1\mathcal{U},0\mathcal{P}}=-c_qc_ts^2_t\varrho_{10,01},\quad\rho_{01,10}=c_qc^3_t\varrho_{01,10},
\nonumber\\
\rho^{(qt)}_{10,10}&=&c^4_t(s^2_q\varrho_{00,00}+c^4_t\varrho_{10,10}),\quad
\rho^{(qt)}_{00,1\mathcal{D}}=c_qc^3_t\varrho_{00,11},
\nonumber\\
\rho^{(qt)}_{1\mathcal{D},1\mathcal{D}}&=&s^2_qc^2_t(s^2_q+\varrho_{01,01})+c^2_t(\varrho_{11,11}+s^2_t\varrho_{10,10}),\quad
\rho^{(qt)}_{0\mathcal{P},1\mathcal{U}}=-c_qc_ts^2_t\varrho_{01,10},\quad
\nonumber\\
\rho^{(qt)}_{1\mathcal{U},1\mathcal{U}}&=&s^2_qc^2_t(\varrho_{02,02}+s^2_t\varrho_{00,00})+c^2_t(\varrho_{12,12}+s^2_t\varrho_{10,10}),
\quad
\rho^{(qt)}_{0\mathcal{U},1\mathcal{P}}=-c_qc_ts^2_t\varrho_{00,11},\quad
\nonumber\\
\rho^{(qt)}_{1\mathcal{P},1\mathcal{P}}&=&s^2_qs^2_t(\varrho_{00,00}s^2+\varrho_{01,01}+\varrho_{02,02})
+s^2_t(s^2_t\varrho_{10,10}+\varrho_{11,11}+\varrho_{11,01}),
\end{eqnarray}
where the suffix $"q"$,  $"t"$  and $"qt"$ refer to the
accelerated qubit, qutrit and qubit-qutrit, respectively.

 The behavior of
entanglement, $\mathcal{E}$ is described in Fig.(1a), where we
assume that both  subsystems are initially polarized on the third
direction, namely we set $s_3=t_3=1$. It is clear that, the
entanglement decreases as the acceleration of the accelerated
subsystems increase. The decay rate of entanglement depends on the
accelerated system. However, if the smaller dimensional system is
accelerated, then the decay rate of entanglement is smaller than
the decay which is caused by accelerating the larger dimensional
system. These results are clearly seen by comparing the behavior
of endamagement when only the qubit is accelerated
(solid-curve)and its behavior when only the qutrit is accelerated
(dash-curve).  Moreover, the decay rate of entanglement is larger
if both subsystems are accelerated.

In Fig.(1b), we consider another system , where we set
$s_2=t_3=0$. The general behavior is similar to that shown in
Fig.(1a), namely, the entanglement decreases as the accelerations
of the subsystem increase. However, the decay rate of entanglement
is larger than that depicted in Fig.(1a).  This shows that, the
  entanglement of the initial system plays an essential
roles on the degree of entanglement of the accelerated systems.

\subsection{ One-Parameter Family}
 This class of qubit-qutrit system is known by  one parameter family. The  density operator of this
 system is given by,

\begin{eqnarray}
\rho_{qt}^{(1)}&=&
\ket{0}_q\bra{0}\Bigl\{\frac{p}{2}\left(\ket{0}_t\bra{0}+\ket{1}_t\bra{1}\right)+\frac{1-2p}{2}\ket{2}_t\bra{2}\Bigr\}
+\ket{1}_q\bra{0}\Bigl\{\frac{p}{2}\ket{2}_t\bra{0}\Bigr\}
\nonumber\\
&+&\ket{0}_q\bra{1}\Bigl\{\frac{p}{2}\ket{0}_t\bra{2}+\frac{1-2p}{2}\ket{2}_t\bra{0}\Bigr\}+
\ket{1}_q\bra{1}\Bigl\{\frac{p}{2}\ket{1}_t\bra{1}+\frac{1-2p}{2}\ket{0}_t\bra{0}\Bigr\},
\end{eqnarray}
where $0\leq p\leq \frac{1}{2}$ and the subscript $"q"$ refers to
the qubit while $"t"$ refers to the qutrit. If we assume that,
only the qubit is accelerated, then the density operator of the
final state in the first region $(I)$ is defined by the  following
non-zero elements, of a matrix of size $2\times 3$,
\begin{eqnarray}
\rho^{(1_q)}_{00,00}&=&\frac{p}{2}c^2_q,\quad
\rho^{(1_q)}_{00,12}=\frac{p}{2}c_q, \quad
\rho^{(1_q)}_{01,01}=\frac{p}{2}c^2_q,\quad
\rho^{(1_q)}_{00,12}=\rho_{12,00},
\nonumber\\
\rho^{(1_q)}_{02,02}&=&\frac{1-2p}{2}c^2_q,\quad\rho^{(1_q)}_{10,02}=\frac{1-2p}{2}c_q,
\quad\rho^{(1_q)}_{10,10}=\left(\frac{1-2p}{2}+\frac{p}{2}s^2_q\right),
\nonumber\\
\rho^{(1_q)}_{02,10}&=&\rho_{10,02},\quad\rho^{(1_q)}_{11,11}=\frac{p}{2}(1+s^2_q),\quad
\rho^{(1_q)}_{12,12}=\left(\frac{1-2p}{2}s^2_q+\frac{p}{2}\right).\quad
\end{eqnarray}
On the other hand, if  only the qutrit is accelerated, then the
final state of the total system  in the first region is defined by
a matrix of size $8\times 8$. The non-zero elements are given by,
\begin{figure}
  \begin{center}
     \includegraphics[width=30pc,height=18pc]{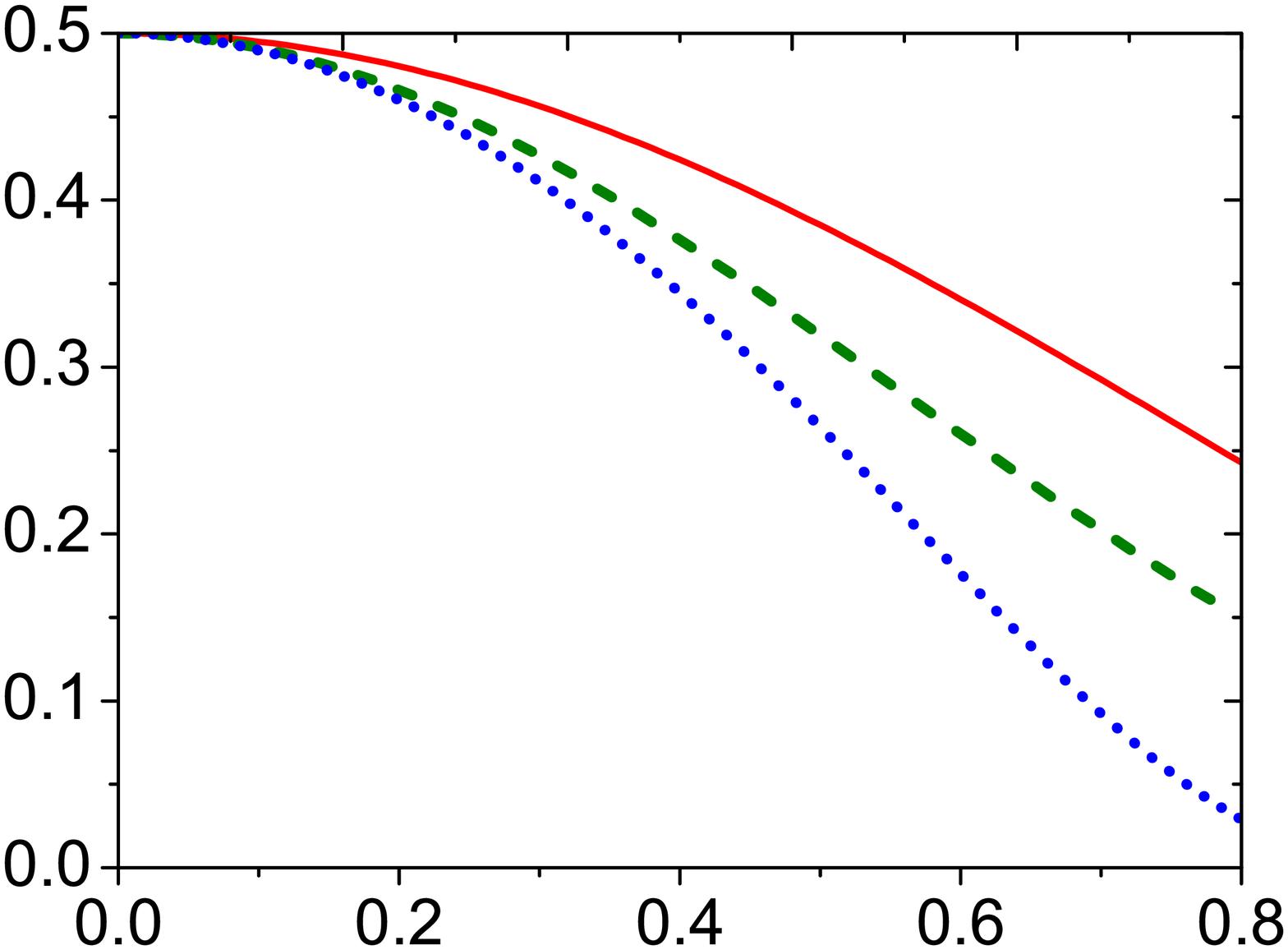}
     \put(-180,0){$r$}
     \put(-335,110){\Large$\mathcal{E}$}
      \caption{ The entanglement of one parameter family evaluated at $p=0.5, r_q=r_t=r$. The solid, dash and dot
       curves represent the entanglement $\mathcal{E}$, when only
       the qubit , only the qutrit, and both of them are accelerated.  }
  \end{center}
\end{figure}
\begin{eqnarray}
\rho^{(1_t)}_{00,00}&=&\frac{p}{2}c^4_t,\quad\rho^{(1_t)}_{12,00}=\frac{p}{2}c^3_t,\quad
\rho^{(1_t)}_{0\mathcal{D},0\mathcal{D}}=\frac{p}{2}c^2_t(1+s^2_t),\quad
\rho^{(1_t)}_{1\mathcal{P},0\mathcal{D}}=\frac{p}{2}c_ts^2_t,
\nonumber\\
\rho^{(1_t)}_{0\mathcal{U},0\mathcal{U}}&=&c^2_t\left(\frac{1-2p}{2}+\frac{p}{2}s^2_t\right),\quad\rho^{(1_t)}_{10,02}=\frac{1-2p}{2}c^3_t,\quad
\rho^{(1_t)}_{0\mathcal{P},0\mathcal{P}}=s^2_t\left(\frac{p}{2}(1+s_t^2)+\frac{1-2p}{2}\right),
\nonumber\\
\rho^{(1_t)}_{1\mathcal{D},0\mathcal{P}}&=&\frac{1-2p}{2}c_ts^2_t,
\quad\rho^{(1_t)}_{10,10}=\frac{1-2p}{2}c^4_t,\quad
\rho^{(1_t)}_{0\mathcal{U},10}=\rho^{(1_t)}_{10,0\mathcal{U}},\quad
\rho^{(1_t)}_{1\mathcal{D},1\mathcal{D}}=\frac{1-2p}{2}c^2_ts^2_t,\quad
\nonumber\\
 \rho^{(1_t)}_{0\mathcal{P},1\mathcal{D}}&=&\frac{1-2p}{2}c_ts^2_t,
 \quad\rho^{(1_t)}_{00,1\mathcal{U}}=\rho^{(1_t)}_{12,00},\quad
 \rho^{(1_t)}_{1\mathcal{U},1\mathcal{U}}=c^2_t\left(\frac{p}{2}+\frac{1-2p}{p}s^2_t\right),
 \nonumber\\
 \rho^{(1_t)}_{0\mathcal{D},1\mathcal{P}}&=&\frac{p}{2}c_ts^2_t, \quad
 \rho^{(1_t)}_{1\mathcal{P},1\mathcal{P}}=s^2_t\left(p+\frac{1-2p}{2}s_t^2\right).
\end{eqnarray}
Finally, if  both subsystems are accelerated, then the final state
in the first region, $I$ is defined by the following non-zero
elements of a matrix of $8\times 8$,
\begin{eqnarray}
\rho^{(1_{qt})}_{00,00}&=&c^2_q\rho^{(1_{t})}_{00,00}, \quad
\rho^{(1_{qt})}_{1\mathcal{D},00}=\frac{p}{2}c_qc^3_t,\quad
\rho^{(1_{qt})}_{0\mathcal{D},0\mathcal{D}}=c^2_q\rho^{(1_{t})}_{0\mathcal{D},0\mathcal{D}},\quad
\rho^{(1_{qt})}_{1\mathcal{P},0\mathcal{D}}=c_q\rho^{(1_{t})}_{1\mathcal{P},0\mathcal{D}},
\nonumber\\
\rho^{(1_{qt})}_{0\mathcal{U},0\mathcal{U}}&=&c^2_q\rho^{(1_{t})}_{0\mathcal{U},0\mathcal{U}},\quad
\rho^{(1_{qt})}_{10,0\mathcal{U}}=c_q\rho^{(1_{t})}_{10,0\mathcal{U}},\quad
\rho^{(1_{qt})}_{0\mathcal{P},0\mathcal{P}}=c^2_q\rho^{(1_{t})}_{0\mathcal{P},0\mathcal{P}},\quad
\rho^{(1_{qt})}_{1\mathcal{D},0\mathcal{P}}=c_q\rho^{(1_{t})}_{1\mathcal{D},0\mathcal{P}},
\nonumber\\
\rho^{(1_{qt})}_{10,10}&=&\rho^{(1_{t})}_{10,10}+\frac{p}{2}c^4_ts^2_q,
\quad
\rho^{(1_{qt})}_{0\mathcal{D},10}=c_q\rho^{(1_{t})}_{10,0\mathcal{U}},\quad
\rho^{(1_{qt})}_{1\mathcal{D},1\mathcal{D}}=\rho^{(1_{t})}_{1\mathcal{D},1\mathcal{D}}+c^2_t\frac{p}{2}\left(1+s^2_q+s^2_qs^2_t\right),
\nonumber\\
\rho^{(1_{qt})}_{0\mathcal{P},1\mathcal{D}}&=&c_q\rho^{(1_{t})}_{0\mathcal{P},1\mathcal{D}},\quad
\rho^{(1_{qt})}_{1\mathcal{U},1\mathcal{U}}=\rho^{(1_{t})}_{1\mathcal{U},1\mathcal{U}}+c^2_t\left(\frac{p}{2}(1+s^2_ts^2_q)+\frac{1-2p}{2}s^2_q\right),\quad
\rho^{(1_{qt})}_{00,1\mathcal{U}}=c_q\rho^{(1_{t})}_{00,1\mathcal{U}},
\nonumber\\
\rho^{(1_{qt})}_{1\mathcal{P},1\mathcal{P}}&=&\rho^{(1_{t})}_{1\mathcal{P},1\mathcal{P}}+s^2_qs^2_t\left(\frac{p}{2}s^2_t+\frac{1-p}{2}\right),\quad
\rho^{(1_{qt})}_{01,1\mathcal{P}}=c_q\rho^{(1_{t})}_{0\mathcal{D},1\mathcal{P}}.
\end{eqnarray}
The behavior of entanglement for this case is shown in Fig.(2),
where we initially start with a system defined by $p=0.5$. The
general behavior is similar to that shown in Fig.(1), namely the
entanglement decays as the the acceleration of the accelerated
subsystem increases. The upper and lower bounds of entanglement
depend on the accelerated subsystem and the initial degree of
entanglement. However, if the qubit is allowed to be accelerated,
then  the upper bounds of entanglement are always larger than that
shown if one allows the qutrit to be accelerated. The decay rate
of entanglement increases if both particles are accelerated, where
we considered  both particles are accelerated with the same
acceleration (i.e., $r_q=r_t=r$).

\subsection { Two-Parameters family}
In this subsection we consider the second example which is known
by a two-parameter family. In the computational basis, this class
can be written as,

\begin{eqnarray}
\rho^{(2)}_{qt}&=&\ket{0}_q\bra{0}\Bigl\{\beta\ket{0}_t\bra{0}+\frac{\beta+\gamma}{2}\ket{1}_t\bra{1}
+\alpha\ket{2}_t\bra{2}\Bigr\}+\ket{1}_q\bra{0}\Bigl\{\frac{\beta-\gamma}{2}\ket{0}_t\bra{1}\Bigr\}
\nonumber\\
&& +
\ket{0}_q\bra{1}\Bigl\{\frac{\beta-\gamma}{2}\ket{1}_t\bra{0}\Bigr\}
+\ket{1}_q\bra{1}\Bigl\{\frac{\beta+\gamma}{2}\ket{0}_t\bra{0}+\beta\ket{1}\bra{1}+\alpha\ket{2}_t\bra{2}\Bigr\},\quad
\end{eqnarray}
where $\gamma+2\alpha+3\beta=1$ and the suffix $2$ refers to the
two-parameter. If it  assumed that only the qubit is accelerated,
then the final state in the first region, $I$ of the Rindler space
is described by the following non-zero-elements of a matrix of
size $6\times 6$,
\begin{eqnarray}
\rho^{(2_{q})}_{00,00}&=&\beta c^2_q,\quad
\rho^{(2_{q})}_{01,01}=\frac{\beta+\gamma}{2}c^2_q,\quad\rho^{(2_{q})}_{10,01}=\frac{\beta-\gamma}{2}c_q,
\nonumber\\
\rho^{(2_{q})}_{02,02}&=&\alpha c^2_q,\quad
\rho^{(2_{q})}_{01,10}=\frac{\beta-\gamma}{2}c_q,\quad
\rho_{10,10}=\beta s^2_q,
\nonumber\\
\rho^{(2_{q})}_{11,11}&=&\left(\beta+\frac{\beta+\gamma}{2}s^2_q\right),\quad
\rho^{(2_{q})}_{12,12}=\alpha(1+s^2_q).
\end{eqnarray}
On the other hand,  if we  consider only the qutrit is accelerated
then, the final state in the first Rindler region is defined by
the following non-zero elements,
\begin{eqnarray}
\rho^{(2_{t})}_{00,00}&=&\beta c^4_t,\quad
\rho^{(2_{t})}_{01,0\mathcal{D}}=c^2_t\left(\beta
s^2_t+\frac{\beta+\gamma}{2}\right),\quad
\rho^{(2_{t})}_{10,0\mathcal{D}}=\frac{\beta-\gamma}{2}c^3_t,
\nonumber\\
\rho^{(2_{t})}_{0\mathcal{U},0\mathcal{U}}&=&c^2_t(\alpha+\beta
s^2_t),\quad
\rho^{(2_{t})}_{0\mathcal{P},0\mathcal{P}}=s^2_t(\left(\alpha+
\frac{3\beta+\gamma}{2}s^2_t\right),\quad
\rho^{(2_{t})}_{1\mathcal{U},0\mathcal{P}}=-\frac{\beta-\gamma}{2}c_ts^2_t,
 \nonumber\\
\rho^{(2_{t})}_{0\mathcal{D},10}&=&\frac{\beta-\gamma}{2}c^3_t,\quad
\rho^{(2_{t})}_{10,10}=\frac{\beta+\gamma}{2}c^4_t,\quad
\rho^{(2_{t})}_{1\mathcal{D},1\mathcal{D}}=c^2_t\left(\beta+\frac{\beta+\gamma}{2}s^2_t\right),
\quad\rho^{(2_{t})}_{0\mathcal{P},1\mathcal{U}}=\rho^{(2_{t})}_{1\mathcal{U},0\mathcal{P}}
\nonumber\\
\rho^{(2_{t})}_{1\mathcal{U},1\mathcal{U}}&=&c^2_t\left(\alpha+\frac{\beta+\gamma}{2}s^2_t\right),\quad
\rho^{(2_{t})}_{1\mathcal{P},1\mathcal{P}}=s^2_t\left(\alpha+\frac{3\beta+\gamma}{2}s^2_t\right).
\end{eqnarray}
Finally, when both subsystems are accelerated, then the final
state in the first region, $I$ is defined by a matrix of size
$8\times 8$ elements. These elements  can be written by (22) as,
\begin{eqnarray} \rho_{00,00}^{(2_{qt})}&=&c^2_q\rho^{(2_{t})}_{00,00},\quad
\rho^{(2_{qt})}_{0\mathcal{D},0\mathcal{D}}=c^2_q\rho^{(2_{t})}_{0\mathcal{D},0\mathcal{D}},
\quad
\rho^{(2_{qt})}_{10,0\mathcal{D}}=c_q\rho^{(2_{t})}_{10,0\mathcal{D}},\quad
\rho^{(2_{qt})}_{0\mathcal{U},0\mathcal{U}}=c^2_q\rho^{(2_{t})}_{0\mathcal{U},0\mathcal{U}}
\nonumber\\
\rho^{(2_{qt})}_{0\mathcal{P},0\mathcal{P}}&=&c^2_q\rho^{(2_{t})}_{0\mathcal{P},0\mathcal{P}},\quad
\rho^{(2_{qt})}_{1\mathcal{U},0\mathcal{P}}=c_q\rho^{(2_{t})}_{1\mathcal{U},0\mathcal{P}},\quad
\rho^{(2_{qt})}_{0\mathcal{D}10}=c_q\rho^{(2_{t})}_{0\mathcal{D},10},\quad
\rho^{(2_{qt})}_{10,10}=s^2_q\rho^{(2_{t})}_{10,10},
\nonumber\\
\rho^{(2_{qt})}_{1\mathcal{D},1\mathcal{D}}&=&\rho^{(2_{t})}_{11,11}+s^2_q\rho^{(2_{t})}_{0\mathcal{D},0\mathcal{D}},
\quad
\rho^{(2_{qt})}_{0\mathcal{P},1\mathcal{U}}=\rho^{(2_{qt})}_{1\mathcal{U},0\mathcal{P}},\quad
\rho^{(2_{qt})}_{1\mathcal{U},1\mathcal{U}}=\rho^{(2_{t})}_{1\mathcal{U},1\mathcal{U}}+s^2_q\rho^{(2_{t})}_{0\mathcal{U},0\mathcal{U}},
\nonumber\\
\rho^{(2_{qt})}_{1p,1p}&=&\rho^{(2_{t})}_{1\mathcal{P},1\mathcal{P}}+s^2_q\rho^{(2_{t})}_{0\mathcal{P},0\mathcal{P}},
\end{eqnarray}

\begin{figure}[t!]
  \begin{center}
     \includegraphics[width=30pc,height=18pc]{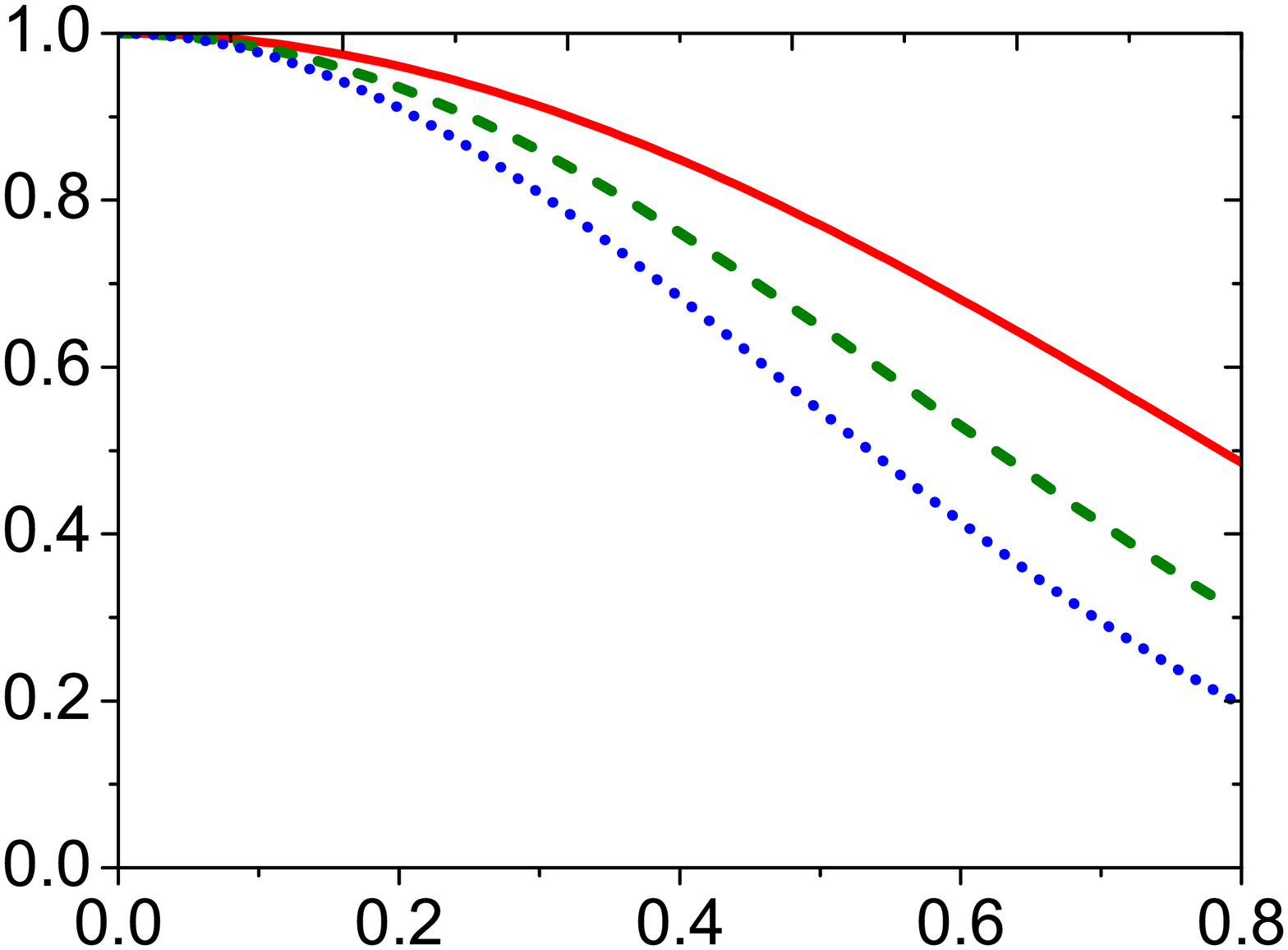}
\put(-180,0){$r$}
  \put(-335,110){\Large$\mathcal{E}$}
      \caption{The dynamics of entanglement in the first region for qubit-qutrit system of two-parameter family
      with $\alpha=\beta=0.5$. The solid, dash and dot-curves represent $\mathcal{E}$, when only the qubit, qutrit
      and both qubit and qutrit are accelerated.    }
  \end{center}
\end{figure}
The behavior of entanglement for this class  is described in
Fig.(3), where all the three possibilities are considered. It is
assumed that, the system is initially prepared in maximum
entangled state $\alpha=\beta=0.5$. Therefore the entanglement is
maximum i.e., $\mathcal{E}=1$ at $r_q=r_t=0$.  It  is clear that,
the behavior of entanglement is similar to that predicated for one
parameter family. However, the entanglement decays as the
acceleration of any subsystem increases, where we consider that,
both of the subsystems have the same acceleration $(r_q=r_t)$. The
rate of entanglement decay increases as the dimensions of the
accelerated  subsystems increase.  Moreover, if both subsystems
are accelerated, then  the entanglement decreases faster.

\section{Conclusion}
In this contribution, we consider a general system  composite of
two different dimensions. One of them is a two-dimensional qubit
and the other is of three dimensional qutrit. It is assumed that,
one or both subsystems are accelerated simultaneously or non-
simultaneously. The density operator of the final state in the
first region  of the accelerated is obtained analytical in a
general form. Different examples are introduced to show the main
task of this manuscript.

Our results show that, the entanglement decays as the acceleration
of the accelerated particle increases. However, the decay rate
depends on the dimensions of the accelerated  subsystem. If only
the qubit is accelerated, then the entanglement decays gradually
to reach its non-zero values at  infinity. On the other hand, if
we allow only  the qutrit to be accelerated, then the entanglement
decays faster and its upper bounds always smaller than that
depicted for accelerating qubit. Moreover, if  one allows for both
subsystems  to be accelerated, then the decay rate of entanglement
is larger.

The idea is shown  explicitly by considering  three different and
common examples. In the first example, we assume that both
subsystems are polarized in $z-$axis. It is clear that, the decay
of entanglement depends on the initial correlation between the two
subsystems. The second and third examples are known by one and
two-parameters families. These examples have been investigated
widely in the context of quantum information. It is shown that the
decay rate of entanglement depends on the initial degree of
entanglement in addition to the dimensions of the accelerated
subsystem.

{\it In conclusion:} the decay of entanglement depends on the
dimensions  of the accelerated subsystems, where the decay rate of
entanglement increases when the larger dimensional  subsystem is
accelerated. The robustness of the accelerated system depends on
the initial quantum correlation between the two subsystems. We
expect that these results could be important in building an
accelerated quantum network consists of different dimensions
nodes.

\appendix
\section*{Appendices}

\section{ The elements of the initial state}
The elements of the initial system  (1) in the basis
$\ket{00},\ket{01},\ket{02},\ket{10},\ket{11}$ and $\ket{12}$ are
given by.
\begin{eqnarray}
\mathcal{A}_1&=&(1+s_3+t_3-c_{33}+t_8/\sqrt{3}-c_{38}/\sqrt{3})/6,\quad
\mathcal{A}_2=(t_1-it_2-c_{31}+ic_{32})/6,
\nonumber\\
\mathcal{A}_3&=&(t_4-it_5-c_{34}+ic_{35})/6,\quad
\mathcal{A}_4=(s_1+is_2+c_{13}+ic_{23}+c_{18}/\sqrt{3}+ic_{28}/\sqrt{3})/6,
\nonumber\\
\mathcal{A}_5&=&(c_{11}-c_{22}+ic_{21}-ic_{12})/6,\quad\mathcal{A}_6=(c_{14}+c_{25}+ic_{24}-ic_{15})/6,
\nonumber\\
\mathcal{A}_7&=&(t_1+it_2-c_{31}-ic_{32})/6, \quad
\mathcal{A}_8=(1-s_3-t_3+t_{8}/\sqrt{3}+c_{33}-c_{38}/\sqrt{3})/6,
\nonumber\\
\mathcal{A}_9&=&(t_6-it_7-c_{36}+ic_{37})/6,\quad
\mathcal{A}_{10}=(c_{11}-c_{22}+ic_{12}-ic_{21})/6,
\nonumber\\
\mathcal{A}_{11}&=&(s_1+is_2-c_{13}+-ic_{23}+c_{18}/\sqrt{3}+ic_{28}/\sqrt{3})/6,
\quad \mathcal{A}_{12}=(c_{16}+c_{27}+ic_{26}-ic_{17})/6,
\nonumber\\
\mathcal{A}_{13}&=&(t_4+it_5-c_{34}-ic_{35})/6, \quad
\mathcal{A}_{14}=(t_6+it_7-c_{36}-ic_{37})/6,
\nonumber\\
\mathcal{A}_{15}&=&(1-s_3-2t_{8}/\sqrt{3}+2c_{38}/\sqrt{3})/6,\quad
\mathcal{A}_{16}=(c_{14}-c_{25}+ic_{15}+ic_{24})/6,
\nonumber\\
\mathcal{A}_{17}&=&(c_{16}-c_{27}+ic_{17}+ic_{26})/6,\quad
\mathcal{A}_{18}=(s_1+is_2-2c_{28}/\sqrt{3}-2c_{18}/\sqrt{3})/6,
\nonumber\\
\mathcal{A}_{19}&=&(s_1+is_2+c_{13}-ic_{23}+c_{18}/\sqrt{3}-ic_{28}/\sqrt{3})/6,\quad
\mathcal{A}_{20}=(c_{11}-c_{22}-ic_{12}-ic_{21})/6,\quad
\nonumber\\
\mathcal{A}_{21}&=&(c_{14}-c_{25}-ic_{15}-ic_{24})/6,\quad
\mathcal{A}_{22}=(1+s_3+t_3+t_{8}/\sqrt{3}+c_{33}+c_{38}/\sqrt{3})/6,
\nonumber\\
\mathcal{A}_{23}&=&(t_1-it_2+c_{31}-ic_{32})/6,\quad
\mathcal{A}_{24}=(t_4-it_5+c_{34}-ic_{35})/6,\quad
\nonumber\\
\mathcal{A}_{25}&=&(c_{11}+c_{22}+ic_{12}-ic_{21})/6,\quad
\mathcal{A}_{26}=(s_1-is_2-c_{13}+ic_{23}+c_{18}/\sqrt{3}-ic_{28}/\sqrt{3})/6,
\nonumber\\
\mathcal{A}_{27}&=&(c_{16}-c_{27}-ic_{17}-ic_{26})/6,\quad
\mathcal{A}_{28}=-(t_1+it_2+c_{31}+ic_{32})/6,
\nonumber\\
\mathcal{A}_{29}&=&(1+s_3-t_3+t_{8}/\sqrt{3}-c_{33}+c_{38}/\sqrt{3})/6,\quad
\mathcal{A}_{30}=(t_6-it_7+c_{36}-ic_{37})/6,\quad
\nonumber\\
\mathcal{A}_{31}&=&(c_{14}+c_{25}+ic_{15}+ic_{24})/6,\quad
\mathcal{A}_{32}=(c_{16}+c_{27}+ic_{17}-ic_{26})/6,
\nonumber\\
\mathcal{A}_{33}&=&(s_1-is_2+2ic_{28}/\sqrt{3}-2c_{18}/\sqrt{3})/6,\quad
\mathcal{A}_{34}=(t_6+it_7+c_{36}+ic_{37})/6,\quad
\nonumber\\
\mathcal{A}_{35}&=&(t_4+it_5+c_{34}+ic_{35})/6,\quad
\mathcal{A}_{36}=(1+s_3-2t_{8}/\sqrt{3}-2c_{38}/\sqrt{3})/6.
\end{eqnarray}

\section{Both subsystems are accelerated}
The parameters $\mathcal{B}_i, i,1..52$ which are appeared in
Eq.(11) are give explicitly as,
\begin{eqnarray}
\mathcal{B}_1&=&c^2_qc^4_t \mathcal{A}_1,\quad
\mathcal{B}_2=c^2_q\mathcal{A}_2,\quad
\mathcal{B}_3=c^2_q\mathcal{A}_3+c_q\mathcal{A}_6,\quad
\mathcal{B}_4=c_qc^4_t\mathcal{A}_4,\quad\mathcal{B}_5=c_qc^3_t\mathcal{A}_5,
\nonumber\\
\mathcal{B}_6&=&c^2_qc^3_t\mathcal{A}_6,\quad\mathcal{B}_7=c^2_qc^2_t(\mathcal{A}_8+s^2_t\mathcal{A}_1),\quad
\mathcal{B}_8=c^qc^2_t\mathcal{A}_9,\quad\mathcal{B}_9=c^2_qc_ts^2_t\mathcal{A}_3,\quad\mathcal{B}_{10}=c_qc^3_t\mathcal{A}_{10}
\nonumber\\
\mathcal{B}_{11}&=&c_qc^2_t(\mathcal{A}_{11}+s^2_t\mathcal{A}_4),\quad\mathcal{B}_{12}=c_qc^2_t\mathcal{A}_{12},\quad
\mathcal{B}_{13}=c_qs^2_t\mathcal{A}_6,\quad\mathcal{B}_{14}=c^2_qc^3_t\mathcal{A}_{13},\quad\mathcal{B}_{15}=c^2_qc^2_t\mathcal{A}_{14}
\nonumber\\
\mathcal{B}_{16}&=&c^qc^2_t(\mathcal{A}_{15}+s^2_t\mathcal{A}_1),\quad
\mathcal{B}_{17}=-c^2_qc_ts^2_t\mathcal{A}_2, \quad
\mathcal{B}_{18}=c_qc^3_t\mathcal{A}_{16},\quad
\mathcal{B}_{19}=c_qc^2_t\mathcal{A}_{19},\quad \nonumber\\
\mathcal{B}_{20}&=&c_q(\mathcal{A}_{18}+c^2_ts^2_t\mathcal{A}_{4}),
\quad\mathcal{B}_{21}=-c_qc^2_ts^2_t\mathcal{A}_5,\quad\mathcal{B}_{22}=c^2_qc_ts^2_t\mathcal{A}_{13},\quad
\mathcal{B}_{23}=-c^2_qc_ts^2_t\mathcal{A}_7, \nonumber\\
\mathcal{B}_{24}&=&c^2_q(s^2_q\mathcal{A}_{15}+s^2_t\mathcal{A}_8+s^4_t\mathcal{A}_1),\quad\mathcal{B}_{25}=c_qc_ts^2_t\mathcal{A}_{16}
,\quad\mathcal{B}_{26}=-c_qc_ts^2_t\mathcal{A}_{10}, \nonumber\\
\mathcal{B}_{27}&=&c_q(s^4_t\mathcal{A}_1+s^2_t\mathcal{A}_{18}+s^2_t\mathcal{A}_{11}),
\quad\mathcal{B}_{28}=c_qc^3_t(c_t\mathcal{A}_{19}+\mathcal{A}_{25}),\quad\mathcal{B}_{29}=c_qc^3_t\mathcal{A}_{20},\quad
\mathcal{B}_{30}=c_qc^3_t\mathcal{A}_{21} \nonumber\\
\mathcal{B}_{31}&=&c^4_t(s^2_q\mathcal{A}_1+\mathcal{A}_{22}),\quad\mathcal{B}_{32}=c^3_t(s^2_q\mathcal{A}_2+\mathcal{A}_{23})
,\quad\mathcal{B}_{33}=c^3_t(s^2_q\mathcal{A}_3+\mathcal{A}_{24})
,\nonumber\\
\mathcal{B}_{34}&=&c_qc^2_t(s^2_q\mathcal{A}_{19}+\mathcal{A}_{26}),\quad
\mathcal{B}_{35}=c_qc^2_t\mathcal{A}_{27},\quad\mathcal{B}_{36}=c_qc_ts^2_t\mathcal{A}_{21},\quad
\mathcal{B}_{37}=c^3_t(s^2_q\mathcal{A}_{7}+\mathcal{A}_{28}),
\nonumber\\
\mathcal{B}_{38}&=&c^2_qs^2_t\mathcal{A}_{22}+c^2_t\mathcal{A}_{29}+s^2_qc^2_t\mathcal{A}_8+s^2_qc^2_ts^2_t\mathcal{A}_1.
\quad \mathcal{B}_{39}=c^2_t(s^2_q\mathcal{A}_9+\mathcal{A}_{30}),
\quad\mathcal{B}_{40}=c_ts^2_t(s^2_q\mathcal{A}_3+\mathcal{A}_{24}),
\nonumber\\
\mathcal{B}_{41}&=&c_qc^3_t\mathcal{A}_{31},\quad\mathcal{B}_{42}=c_qc^2_t\mathcal{A}_{32},\quad
\mathcal{B}_{43}=c_qc^2_t(s^2_t\mathcal{A}_{19}+\mathcal{A}_{33}),\quad
\mathcal{B}_{44}=-c_qc_ts^2_t\mathcal{A}_{20},
\nonumber\\
\mathcal{B}_{45}&=&c^3_t(s^2_q\mathcal{a}_{13}+\mathcal{A}_{34}),\quad
\mathcal{B}_{46}=c_t(c_t\mathcal{A}_{35}+s^2_t\mathcal{A}_{14})
,\quad\mathcal{B}_{47}=c^2s^2\mathcal{A}_{22}+c^2\mathcal{A}_{36}+s^2_qc^2_t\mathcal{A}_{15}+s^2_qc^2_ts^2_t\mathcal{A}_1
\nonumber\\
\mathcal{B}_{48}&=&c_qc_ts^2_q(\mathcal{A}_{31}-\mathcal{A}_{25}),
\quad\mathcal{B}_{49}=c_qs^2_t(s^2\mathcal{A}_{19}+s^2_q\mathcal{A}_{26}+\mathcal{A}_{33}),\quad
\mathcal{B}_{50}=c_ts^2_t(s^2q\mathcal{A}_{13}+\mathcal{A}_{34}),
\nonumber\\
\mathcal{B}_{51}&=&-c_ts^2_t(s^2_q\mathcal{A}_7+\mathcal{A}_{28}),\quad
\mathcal{B}_{52}=s^2\mathcal{A}_{29}+s^2_qs^2_t(\mathcal{A}_8+\mathcal{A}_{15})+s^2_qs^4_t\mathcal{A}_1+s^3_t\mathcal{A}_{22}+
s^2\mathcal{A}_{36}.
\end{eqnarray}

\end{document}